\documentclass[preprint,12pt]{elsarticle}

\usepackage{amssymb}
\usepackage{amsmath}
\usepackage{booktabs}
\usepackage{xcolor}
\usepackage{tabularray}
\usepackage{dirtree}
\usepackage{url}
\usepackage[frozencache]{minted}

\newcounter{bla}

 \newcommand{\repourl}{\url{https://github.com/qmolastro/madwave3}}

\journal{Computer Physics Communications}

\begin{document}

\begin{frontmatter}



\title{MADWAVE3: a quantum time dependent wave packet code for
    nonadiabatic state-to-state reaction dynamics  of triatomic
    systems .}

\author[a]{Octavio Roncero\corref{author}}
\author[b]{Pablo del Mazo-Sevillano}

\cortext[author] {Corresponding author.\\\textit{E-mail address:} octavio.roncero@csic.es}
\address[a]{Instituto de F{\'\i}sica Fundamental (IFF-CSIC), Serrano 123, 28006 Madrid (Spain)}
\address[b]{Unidad Asociada UAM-IFF-CSIC, Departamento de  Qu{\'\i}mica F{\'\i}sica Aplicada, Facultad de Ciencias M-14, Universidad Aut\'onoma de Madrid, 28049 Madrid (Spain)}

\begin{abstract}
We present MADWAVE3, a FORTRAN90 code designed for quantum time-dependent wave packet propagation in triatomic systems. This program allows the calculation of state-to-state probabilities for inelastic and reactive collisions, as well as photodissociation processes, over one or multiple coupled diabatic electronic states. The code is highly parallelized using MPI and OpenMP. The execution requires the potential energy surfaces of the different electronic states involved, as well as the transition dipole moments for photodissociation processes. The formalism underlying the code is presented in  section 2, together with the modular structure of the code. This is followed by the installation procedures and  a comprehensive list and explanation of the parameters that control the code, organized within their respective namelists.

Finally, a case study is presented, focusing on the prototypical reactive collision H+DH($v,j$)$\rightarrow$ H$_2$($v',j'$) + D. Both the potential energy surface and the input files required to reproduce the calculation are provided and are available on the repository’s main page. { This example is used to study the parallelization speedup of the code.}

{\bf NEW VERSION PROGRAM SUMMARY}

\begin{small}
\noindent
{\em Program Title:} MADWAVE3                                         \\
{\em CPC Library link to program files:} (to be added by Technical Editor) \\
{\em Developer's repository link:} \repourl \\
{\em Code Ocean capsule:} (to be added by Technical Editor)\\
{\em Licensing provisions(please choose one):} GPLv3\\
{\em Programming language:} Fortran 90     
{\em External libraries:} FFTW3, MPI\\
{\em Supplementary material:}                                 \\
{\em Nature of problem:} Quantum time propagation of a wave packet describing a reactive process in a triatomic systems, for collisions (inelastic and reactive) and photodissociation processes, and considering several coupled diabatic electronic state\\
{\em Solution method(approx. 50-250 words):} A modified Chebyshev propagator is used, keeping the real Chebyshev components, which are represented in grids for the internal Jacobi coordinates  $r, R$ and $\gamma$ and in a basis for electronic and helicity components. The potential represented in a grid as well as the reactants and products wave functions are previously calculated in a preparatory stage.\\
{\em Additional comments including restrictions and unusual features (approx. 50-250 words):}\\
   \\

\end{small}


\end{abstract}


\end{frontmatter}


\section{Introduction}\label{sec:section}

MADWAVE3 is a code written in FORTRAN90 to treat quantum state-to-state reaction dynamics for A+BC(v,j) $\rightarrow$ AB(v',j')+C collisions and ABC+h$\nu$ $\rightarrow$ AB(v',j')+C photodissociation processes. The code is parallelized using MPI and OpenMP, and this is done primarily on the helicity quantum number, as described below~\cite{Roncero-etal:97}. Non-adiabatic processes are also
included by considering multiple electronic states in a diabatic representation, in which the couplings are only due to potential terms, and kinetic coupling terms are neglected. 

MADWAVE3 comes with a set of helper programs in order to prepare the calculations, such as generating the potential energy surface grids, and to analyze the results, such as calculating the process cross sections.

To extract the state-to-state probabilities, the Jacobi coordinates of the reactants or products can be used, depending on the masses involved and the symmetry of the reaction~\cite{Gomez-Carrasco-Roncero:06}. In the reactant Jacobi coordinate based (RCB) method, the wave packet is transformed from the coordinates of the reactants to the coordinates of the products in a sequential procedure at each iteration,
in order to reduce computational time~\cite{Gomez-Carrasco-Roncero:06}. When using products' coordinates, the transformation is done only once for the initial wave packet.

{ One popular alternative to run reactive collisions is the ABC code~\cite{Skouteris-etal:00}. ABC is a time-independent close-coupling code using hyperspherical coordinates for A+BC reactive collisions, and is easier to converge at low energies and for narrow resonances. However, it scales as $n^2$ in memory and as $n^3$ in computing time with the number of channels, $n$. This is the case for high angular momentum requiring high helicities and for deep insertion wells, where the basis formed by all rearrangement products is sometimes insufficient to converge the calculations. In such a situation, MADWAVE3 is an interesting alternative which scales better with the number of channels since it deals with the calculation of one column of the whole S-matrix.  Moreover, MADWAVE3 is highly parallelizable, as commented below, and considers the case of multiple coupled electronic states and can treat photodissociation dynamics, not implemented in the ABC code.

A very popular code for studying photodissociation processes is the time-dependent MCTDH\cite{Meyer-etal:90,Manthe-etal:92,Beck-etal:00} program, which scales very well
with increasing system size, since it uses mono-dimensional functions to expand the full wave functions. This allows accurate calculations of the photodissociation cross section, for large systems and including multiple coupled electronic states. However, when the wave packet splits in many portions it is very difficult to converge the basis formed by products of monodimensional functions and there are only few calculations of reactive collisions using MCTDH code. 

An alternative is the use of the MCTDH approach within the framework of the thermal ﬂux operator \cite{Manthe-etal:93,Huarte-Larranaga-Manthe:00} to calculate thermal rate constants in reactions involving polyatomic molecules. This method has been extended to treat state-to-state reaction dynamics in polyatomic systems \cite{Welsch-Manthe:15} showing their power. However, this approach is limited to reactions with barriers. In this regard, MADWAVE3 can treat reactions without barriers but is limited to reactions with three atoms or using reduced dimension models for larger systems.

The DIFFREALWAVE\cite{Hankel-etal:06} code is similar to MADWAVE3, and it is designed to calculate state-to-state differential cross sections in atom plus diatom reactive collisions. This program uses product Jacobi coordinates and one single electronic state. This code is not designed for the study of photodissociation processes, while MADWAVE3 is.

The use of hyperspherical coordinates in wave packet calculations \cite{Ghosh-etal:15,Ghosh-etal:18} is an interesting alternative to state-to-state dynamics. The main problem with this approach is the difficulty of describing long-distance regions on each rearrangement channel, which requires very dense grids and basis in the hyperangles to describe the rovibrational states of diatomic fragments. This limits the use of this approach to relatively high energies.

Other related methods are the reactant-product decoupling (RPD) \cite{Althorpe:01,Peng-Zhang:96,Zhu-etal:97,Dai-Zhang:97}, in which the wave packet is split into reactant and product channels, and its evolution is calculated separately in the most appropriate, while the flux between the different channels is exchanged using negative imaginary potentials. However, this method shows difficulties when the reaction presents wells giving rise to resonances, which are affected by the split description and the imaginary potential. This problem has recently been addressed \cite{Shu-etal:24} opening for new applications in the future.
}

In the following, we start introducing the theory behind the MADWAVE3 program in section~\ref{sec:theory},
and its implementation in modules of the code. {Next, in section~\ref{sec:distribution}  the installation procedures are described. In section~\ref{sec:dependent_programs}, the system dependent programs needed to be applied to particular systems are described. Section~\ref{sec:input_data}, in  Table~\ref{tab:parameters} provides a comprehensive description of the various FORTRAN namelists and parameters involved in the code execution.
Section~\ref{sec:example} shows  a benchmark  example with steps to follow the execution. The parallelization efficiency of the code is presented in section~\ref{sec:parallelization}, using the example of section~\ref{sec:example}. Finally, a summary is presented in the conclusions.}

\section{Theory}\label{sec:theory}
\subsection{Coordinates and Hamiltonian}

To describe the A+BC channel the Jacobi vectors ${\bf r}_{\alpha=A}$ (BC inter-nuclear vector) and
${\bf R}_{\alpha=A}$ (the vector joining BC center-of-mass to A atom) are defined. There are three sets of these coordinates, $\alpha=$ A, B and C defined in a cyclic way, each one adapted to describe the corresponding asymptotic solution as a simple product of monodimensional eigenfunctions for each rearrangement. 
Hereafter, the subscript $\alpha$ will be generally omitted (unless necessary) since the equations are formally equivalent for each set.

A body-fixed frame is used, in which the z-axis is parallel to the ${\bf R}$ vector, and ${\bf r}$ is in the x-z body-fixed frame. The body-fixed is obtained by a rotation, defined by the Euler angles ($\phi,\theta,\chi$),
from the space-fixed frame, following the convention defined by Ref.~\cite{Zare-book}. 
Thus, in the body-fixed frame, the coordinates are divided into the three external Euler angles and the three internal coordinates $(r, R, \gamma)$, where $r$ and $R$ are the magnitudes of vectors $\mathbf{r}$ and $\mathbf{R}$, respectively, and $\gamma = \mathbf{r} \cdot \mathbf{R} / {r R}$ represents the angle between them.
  In these coordinates, the total wave packet is expressed as
\begin{eqnarray}\label{eq:total-function}
    \Psi^{JMp}_i({\bf r},{\bf R};t)= \sum_{\Omega\Lambda} \frac{\Phi^{JMp}_{\Omega\Lambda}(r,R,\gamma;t)}{rR}
    W^{JMp}_{\Omega\Lambda}(\phi,\theta,\chi),
\end{eqnarray}
where $i$ is a collection of quantum numbers that specify the initial state, and the angular-electronic functions are
\begin{eqnarray}\label{eq:angular-functions}
    W^{JMp}_{\Omega\Lambda}(\phi,\theta,\chi)=\sqrt{ \frac{2J+1}{16\pi^2 (1+\delta_{\Omega0})}} &&\Bigg\lbrack D^{J*}_{M\Omega}( \phi,\theta,\chi) \,\vert\Lambda,\sigma\rangle \\&+&  p \sigma(-1)^{J+\Omega}D^{J*}_{M-\Omega}( \phi,\theta,\chi) \vert -\Lambda,\sigma\rangle \ \Bigg\rbrack \nonumber
\end{eqnarray}
with $J,M,\Omega$ being the total angular momentum, and its projections
onto the space-fixed and body-fixed z-axis, respectively, $p$ is the parity under inversion of spatial coordinates. $D^{J*}_{M\Omega}( \phi,\theta,\chi)$ are
Wigner rotation matrices~\cite{Zare-book} and $\vert\Lambda,\sigma\rangle$ are electronic functions, with $\sigma$ being the parity under reflection through the x-z body-fixed plane.

The $\Phi^{JMp}_{\Omega\Lambda}(r,R,\gamma;t)$ coefficients in Eq.~(\ref{eq:total-function}) 
are represented in grids as
\begin{eqnarray}
\Phi^{JMp}_{\Omega\Lambda}(r,R,\gamma;t) = \sum_{k,m,n}
\langle r_k R_m \gamma_n\vert \Phi^{JMp}_{\Omega\Lambda}(t)\rangle ,
\end{eqnarray}
using Gauss-Legendre quadrature points for $\gamma$, and equidistant points for the radial coordinates, $r, R$. The grids and basis set in reactants Jacobi coordinates are controlled through the {\bf inputgridbase} namelist (see Table~\ref{tab:parameters}).

The triatomic Hamiltonian in Jacobi coordinates is written as
\begin{eqnarray}\label{eq:hamiltonian}
    H= -\frac{\hbar^2}{2\mu}\left( \frac{2}{R}\frac{\partial}{\partial R}+\frac{\partial ^2}{\partial R^2}\right)-\frac{\hbar^2}{2m}\left( \frac{2}{r}\frac{\partial}{\partial r}+\frac{\partial^2}{\partial r^2}\right)
    +\frac{\hat{\ell}^2}{2\mu R^2}+\frac{\hat{j}^2}{2 m r^2}
    +H_{e}
\end{eqnarray}
with $\mu=m_A(m_B+m_C)/(m_A+m_B+m_C)$, $m=m_Bm_C/(m_B+m_C)$ being the reduced masses, $\hat{j}$ and $\hat{\ell}$ are the angular momentum operators
associated to ${\bf r}$ and ${\bf R}$. In the body-fixed frame
$\hat{\ell}^2=\hat{J}^2 + \hat{j}^2 - 2 \hat{J}\cdot\hat{j}$. Finally,
$H_e$ is the electronic Hamiltonian represented in a diabatic electronic basis set $\vert\Lambda,\sigma\rangle$, so that
$\left\langle \Lambda,\sigma\vert H_e\vert \Lambda',\sigma'\right\rangle = f(r,R,\gamma)$ are simple potential terms (with no kinetic couplings). Hereafter, it will be considered that asymptotically (when $R\rightarrow\infty$) the electronic matrix is diagonal for one rearrangement (typically the one of the reactants) but not necessarily for the other two.

Insertion of Eq.~(\ref{eq:total-function}) into the time-dependent Schr\"odinger equation using the Hamiltonian of Eq.~(\ref{eq:hamiltonian}), yields a set of first order differential equations for the $\langle r_k R_m \gamma_n\vert \Phi^{JMp}_{\Omega\Lambda}(t)\rangle$ coefficients
\begin{eqnarray}
    i\hbar{\partial\langle r_k R_m \gamma_n\vert \Phi^{JMp}_{\Omega\Lambda}(t)\rangle\over \partial t} &=& \sum_{k'm'n'\Omega'\Lambda'} \Bigg\lbrace \delta_{mm'}\delta_{nn'}\delta_{\Omega\Omega'}\delta_{\Lambda\Lambda'} 
    \left\langle r_k \left\vert -{\hbar^2\over 2 m}{\partial ^2 \over\partial r^2}\right\vert r_{k'}\right\rangle \nonumber \\
 &+& \delta_{kk'}\delta_{nn'}\delta_{\Omega\Omega'}\delta_{\Lambda\Lambda'} 
    \left\langle R_m \left\vert -{\hbar^2\over 2\mu}{\partial ^2 \over\partial R^2}\right\vert R_{m'}\right\rangle \nonumber \\    
 &+& \delta_{kk'}\delta_{mm'}\delta_{\Omega\Omega'}\delta_{\Lambda\Lambda'} 
    \left\langle \gamma_n \left\vert {\hat{j}^2\over 2 m r_k^2}\right\vert \gamma_{n'}\right\rangle  \\    
 &+& \delta_{kk'}\delta_{mm'}\delta_{\Lambda\Lambda'} 
    \left\langle \  \gamma_n  \left\vert {
    \langle W^{JMp}_{\Omega\Lambda}\vert\hat{\ell}^2\vert W^{JMp}_{\Omega'\Lambda}\rangle
    \over 2\mu R^2}\right\vert \gamma_{n'}\right\rangle \nonumber \\    
 &+& \delta_{kk'}\delta_{mm'}\delta_{nn'}\delta_{\Omega\Omega'}
    \left\langle \Lambda,\sigma \left\vert H_e\right\vert \Lambda',\sigma'\right\rangle\Bigg\rbrace \langle r_{k'} R_{m'} \gamma_{n'}\vert \Phi^{JMp}_{\Omega'\Lambda'}(t)\rangle\nonumber   
\end{eqnarray}

The $\langle r_k R_m \gamma_n\vert \Phi^{JMp}_{\Omega\Lambda}(t)\rangle$ coefficients depend on 5 indexes: $k,m,n,\Omega,\Lambda$, and each term of the Hamiltonian is non-diagonal in only one variable, except $\hat{\ell}^2$, which is non-diagonal in two: $n$ and $\Omega$. However, the Coriolis term  only couples $\Omega$ values  with $\Delta \Omega=0 , \pm 1$, {\it i.e.} it is a tridiagonal term. All this allows to asses:
\begin{enumerate}
    \item Each term of the Hamiltonian is applied sequentially, which means that the multiplication of a monodimensional matrix and a vector, within the 5 dimensional vector of coefficients. This implies 6 nested loops.
    
    \item The two radial terms are further accelerated by applying a 
    Fourier transform method using the FFTW3 library\footnote{The FFTW3 library is not thread-safe and for some compilers it may result in wrong executions. Please be aware of this problem and compare with the executions compiled with the {\bf -fno-omp} flag in gfortran compiler.} (the current version is not ready for the version included in MKL library). In particular, we use the real sine transform to impose the proper regular boundary conditions at $R=0$ \cite{Lepetit-Lemoine:02,Gonzalez-Lezana-etal:05}, and to keep the wave packet propagation real, as will be described below. 

    \item The two angular kinetic terms, $\hat{j}^2$ and $\hat{\ell}^2$, are applied following a transformation from a discrete variable representation (DVR) to a finite basis set (FBR), in which the matrix elements are analytical, as explained in Ref.~\cite{Roncero-etal:97}.
    
    \item The $\hat{\ell}^2$ term, presenting the extra difficulty of being tridiagonal in $\Omega$ is easily parallelized in this index, thus minimizing the communication among processors~\cite{Roncero-etal:97}.

    \item The electronic Hamiltonian (hereafter potential term), is diagonal in the internal coordinates. This term can reach very high values, so it is generally advisable to limit it to a finite maximum value, out of the physically relevant interval. This approach also significantly reduces memory usage, as the entire configuration space can be limited to only the accessible region, known as L-shaped grids~\cite{Mowrey:91}. This reduction is done before running the wave packet calculations, in the preparation of the potential matrix elements, read by the MADWAVE3 program.
\end{enumerate}

The action of the Hamiltonian on the wave packet is evaluated in module {\bf mod\_Hphi\_01y2.f}.
The potential is read from the directory {\bf ../pot}, which is generated in the preparation stage 
with the system dependent code {\bf pot.out} (see section~\ref{sec:dependent_programs}) using the module {\bf  mod\_pot\_01y2.f}
on the grid defined above, and the {\bf vcutmax\_eV} in {\bf namelist inputpotmass}
where also the atomic masses are provided. The grid and basis sets
are defined in modules {\bf mod\_gridYpara\_01y2.f} and {\bf mod\_baseYfunciones\_01y2.f}

\subsection{Propagation and absorption}

The propagation is performed with a modified Chebyshev propagator, proposed by Mandelshtam and Taylor~\cite{Mandelshtam-Taylor:95} as a modification including absorption of the original Chebyshev propagator by Tal-Ezer and Kosloff~\cite{TalEzer-Kosloff:84}. Several alternative but equivalent forms of these propagators have been proposed~\cite{Huang-etal:94,Huang-etal:96,Kroes-Neuhauser:96,Chen-Guo:96,Gray-Balint-Kurti:98}. Basically, the Chebyshev components of the wave packet are obtained iteratively as 
\begin{eqnarray}\label{eq:wvp-k}
\Psi(k=0)&=&\Psi(t=0) \nonumber \\
\Psi(k=1)&=& e^{-\varphi}{\hat H}_s \Psi(k=0)\\
\Psi(k+1)&=&e^{-\varphi} \left\lbrace 2 {\hat H}_s\Psi(k) -e^{-\varphi}\Psi(k-1)
             \right\rbrace,
             \nonumber
\end{eqnarray}
where ${\hat H}_s= \left( {\hat H} -E_0 \right)/ \Delta E$ is the scaled Hamiltonian, with
 $E_0= (E_{max}+E_{min})/2$, $\Delta E=
(E_{max}-E_{min})/2$ and $E_s=(E-E_0)/\Delta E$, $E_{max}$ and $E_{min}$ being the minimum and
 maximum energy values of the Hamiltonian of the system represented
 in the grid/basis used in the propagation. In Eq.~\eqref{eq:wvp-k}, the initial Chebyshev component is defined as the initial wave packet, and $e^{-\varphi}$ is a real damping function, introduced
 to avoid transmission and reflection of the wave packet in finite grids when the edges are reached.
 To first order, 
 the $\varphi$ function can be related to a purely imaginary
potential, $V_I$ ($V_I\approx i \varphi$), fulfilling
\begin{eqnarray}\label{eq:imaginary-potential}
\varphi(x)= \left\lbrace
\begin{array}{ccc}
0 & {\rm for }& x<x_I \\
A_x (x-x_I)^{n_I} & {\rm for} & x > x_I
\end{array}
\right. ,
\end{eqnarray}
with $x\equiv r$ or $R$. 
The absorption parameters $n_I$,$x_I$ ($R_I, r_I$) and $A_x$
($A_R, A_r$) need to be optimized to minimize
the absorption effects on the wave packet in the interaction 
region~\cite{Riss-Meyer:96}. The transmission through the absorption
region, $i.e.$ from $x_I$ to $x_{max}$, can be strongly reduced by
increasing either $A_x$ or $x_{max}$. The major problem, however,
is the reduction of the reflection back to the interaction
region~\cite{Riss-Meyer:96}. This is the case when low final
translation energies between reactants or products play an important
role, in the opening of new channels.
 When the de Broglie wavelength, $\lambda_{dB}$,
 increases, the absorption region should be augmented to an integer
 factor of $\lambda_{dB}$, but this may lead in some cases to use 
 prohibitively large grids.

With these definitions, the Chebyshev components stay real if the initial wave packet 
is real, what allows to considerably reduce the memory requirements in computational implementations. Also, working in the Chebyshev domain allows to reduce the number of iterations
required to get magnitudes defined either in the time- or the energy-domain. This is achieved by transforming either to a time-dependent wave packet or to an energy-dependent wave function using the expressions 
\begin{eqnarray}\label{eq:wvf-chebyshev}
\Psi(t)&=& 
\sum_{k=0}^\infty f_k({\hat H}_s,t) \Psi(k) \nonumber \\
\Psi(E)&=& {1\over a_0(E)}\sum_{k=0}^\infty c_k({\hat H}_s,E) \Psi(k)
\end{eqnarray}
where $a_0(E)$ is defined below, and 
\begin{eqnarray}\label{eq:chebyshev-expansion-constants}
 f_k({\hat H}_s,t) &=& \left( 2-\delta_{k0}\right)  e^{-i E_0 t /\hbar}
 \,(-i)^k   \, J_k( t \Delta E /\hbar)\\
  c_k({\hat H}_s,E) &=& \left( 2-\delta_{k0}\right) {\hbar \exp\left\lbrack-i k \arccos E_s \right\rbrack\over \sqrt{\Delta E^2 -(E-E_0)^2}}
\nonumber                        
\end{eqnarray}
with $J_k$ being Bessel functions of the first kind. 

The Chebyshev propagation is performed in the main program {\bf main\_madwave3.f} using the input parameters defined in {\bf namelist inputtime} (in input.dat). The absorption 
is defined in module {\bf mod\_absorcion\_01y2.f } using the parameters provided by the user in the {\bf namelist inputabsorcion} (in input.dat).

\subsection{Initial wave packet}

Two cases will be considered: one for collisions on the reactants channel, and the other 
for photodissociation in the interaction region. These are defined separately below. The process is defined by the variable {\bf iphoto} = 0 or 1 in the {\bf namelist inputprocess} (in input.dat) and is read in module 
{\bf mod\_gridYpara\_01y2.f}

\subsubsection{Initial wave packet in collisions}

The initial wave packet for collisions is defined for a particular rovibrational
state of the reactants, given by the quantum number $\alpha=$ $J$, $p$, $\Lambda_0$, $v_0,j_0,\Omega_0$, as
\begin{eqnarray}
\label{eq:initial-collision-wvp}
\Psi^\alpha(t=0)= W^{JMp}_{\Omega_0\Lambda_0} \chi_{v_0j_0}(r)
Y_{j_0\Omega_0}(\gamma) g(R),
\end{eqnarray}
where $\chi_{vj}(r)$ are the vibrational eigenfunctions of the diatomic
reactant, $Y_{j\Omega}(\gamma)$ are normalized associated Legendre functions
and $g(R)$ is a real function obtained as a linear superposition of 
incoming and outgoing plane
waves of the form \cite{Gray-Balint-Kurti:98}:
\begin{eqnarray}\label{eq:initial-gaussian}
g(R) =  {e^{-(R-R_0)^2/2\Gamma^2}\over
 2 \left\lbrack\pi \Gamma^2\right\rbrack^{1/4}} 
\left\lbrack e^{-ik_0 R} + e^{ik_0 R} \right\rbrack.
\end{eqnarray} 
As mentioned in Ref.~\cite{Gray-Balint-Kurti:98}
only the incoming half of the initial wave packet yields to
products. Therefore, the evaluation of the reaction probabilities 
 only requires the initial flux 
corresponding to the incoming waves, $i.e.$, at a
given energy, only a half of the flux is taken into account.

The incoming flux in the initial state of the reactants, provided
that the wave packet at $t=0$ is located at a sufficiently long distance
where the potential is constant, is defined as
\begin{eqnarray}\label{initial-flux-collisions}
    a_0(E)= {1\over 2i} \sqrt{ {\mu\over 2\pi \hbar^2 K_0(E)} } \int dR\, e^{iK_0(E)R}\, g(R)
\end{eqnarray}
with $K_0(E)= \sqrt{2 \mu (E-E_{v0j0})/\hbar^2}$, with $E_{v0j0}$ being the energy of the initial rovibrational eigenstate of the BC reagent.

The initial wave packet in collisions is defined in the module {\bf mod\_colini\_01y2.f}
using the data in {\bf namelist inputcol}.
    
\subsubsection{Initial wave packet in photodissociation}

In the framework of first order perturbation theory for the photodissociation of an initial bound state, $\Psi^{J_iM_ip_i}_{X i}$, in electronic state X and vibrorotational state $i$, to a final electronic state $\Lambda$ in a particular rotational state $J$, the initial wave packet is built as~\cite{Paniagua-etal:99}
\begin{eqnarray}\label{eq:initial-photo-wvp}
   \left\vert\Psi^{JMp\leftarrow J_iM_ip_i}_{\Lambda}(t=0)\right\rangle =\sum_\Omega \left\vert W^{JMp}_{\Omega_\Lambda}\right\rangle 
   \left\langle W^{JMp}_{\Omega_\Lambda} \vert {\bf d}\cdot{\bf e}\vert \Psi^{J_iM_ip_i}_{X i}\right\rangle,
\end{eqnarray}
where ${\bf d}$ are the matrix elements of the electric dipole moment between the ground X and the excited $\Lambda$ electronic states, which depends on the internal coordinates $r, R, \gamma$. Its components are expressed in the molecular body-fixed frame, while the polarization vector, ${\bf e}$, of the incident photon defines the space fixed frame, so that the transition operator is written as
\begin{eqnarray}
    {\bf d}\cdot{\bf e}= \sum_ {pq} (-1)^p ({\bf e})_{-p} \, D^{1*}_{pq}(\phi,\theta,\chi) \, d_q (r,R,\gamma),
\end{eqnarray}
where $p=0$ for linearly polarized light (with the z space-fixed frame along the polarization vector of the photon) or $p=\pm 1$
for left/right circularly polarized (with the z space-fixed axis along the propagation of the incident photon).

The energy content of the initial wave packet is not known in this case,
and needs to be evaluated along the propagation. This is done by doing the Fourier transform
\begin{eqnarray}\label{eq:fourier-transform}
    a_0(E)= { 1\over 2\pi\hbar} \int_{-\infty}^\infty dt \, e^{iEt\hbar}
    \,\left\langle \Psi^{JMp\leftarrow J_iM_ip_i}_\Lambda(t=0)\bigg\vert \Psi^{JMp\leftarrow J_iM_ip_i}_\Lambda(t)\right\rangle 
\end{eqnarray}
which using Eqs.~\eqref{eq:chebyshev-expansion-constants}, in terms
of the Chebyshev components becomes
 \begin{eqnarray}\label{eq:chebyshev-transform}
    a_0(E)= { 1\over \pi} \sum_{k=0}^\infty (2-\delta_{k0}) \,
    { \cos(-k \arccos E_s)\over \Delta E\sqrt{1-E_s^2}}
    \,\left\langle \Psi^{JMp\leftarrow J_iM_ip_i}_\Lambda(k=0)\bigg\vert \Psi^{JMp\leftarrow J_iM_ip_i}_\Lambda(k)\right\rangle 
\end{eqnarray}

The initial wave packet for photodissociation processes is defined in module {\bf mod\_photoini\_01y2.f}, using the data defined in {\bf namelist inputbnd}. It reads
system dependent data: the initial bound state in directory {\bf ../bnd} and the transition dipole moment in directory {\bf ../dip}. The initial bound state can be calculated using 
the {\bf main\_boundlanz.f } program (also in this version with compiled name {\bf bound.out})
which calculates the bound states using an iterative non-orthogonal Lanczos method \cite{Cullum-Willoughby:85} in a grid, and the matrix elements are evaluated as described above for the wave packet. The transition dipole moment is evaluated in the preparation stage, by the main program {\bf pot.out}, calling the routine {\bf dipele.f} which should be supplied for each system as described below.

\subsection{Analysis of final states}

When the final state distribution is not needed, and only the reaction probability versus energy is required, the method of the total flux
is probably the best choice~\cite{Miller:74,Zhang-Zhang:94a,Neuhauser:94,Aguado-etal:97b}. The total reaction probability then becomes
\begin{eqnarray}\label{eq:total-flux}
    P^{Jp\alpha}(E) &=& \sum_{\beta v' j'\Omega'} \left\vert S^{Jp}_{v_0j_0\Omega_0, \beta v',j'\Omega'} (E)\right\vert^2\\
    &=& {K_0(E)\over \mu m} \sum_\Omega\sum_\Lambda \int dR \sin\gamma d\gamma
     Im\left\lbrack  \Psi^{Jp\alpha + *}(E){\partial \Psi^{Jp\alpha + *}(E) \over \partial r} \Bigg\vert_{r^*}\right\rbrack \nonumber,
\end{eqnarray}
where $\Psi^{Jp\alpha + *}(E)$ is given in Eq.~\eqref{eq:wvp-k} and $r^*$ is a sufficiently long BC internuclear distance. Since positive outgoing and negative incoming flux are accounted for, the total reaction probability is independent 
of the particular value of $r^*$~\cite{Miller:74}. Thus $r^*$ does not need 
to be in the asymptotic region; it is typically positioned just beyond the saddle point for the reaction, where the flux converges more quickly.

The total flux is evaluated in module {\bf mod\_flux\_01y2.f} using the parameters defined in {\bf namelist inputflux} (in input.dat). 

In Eq.~\eqref{eq:total-flux}, the summation runs over the products rearrangements, $\beta=B, C$, and all final rovibrational states of the products, $v',j',\Omega'$, and $S^{Jp}_{v_0j_0\Omega_0\Lambda_0, \beta v',j'\Omega'} (E)$ are matrix elements of the scattering S-matrix. The flux for individual final states is determined by the square of the corresponding S-matrix element, $\left\vert S^{Jp}_{v_0j_0\Omega_0 \Lambda_0, \beta v',j'\Omega'} (E)\right\vert^2$ and need to be computed at sufficiently long distances so that the interaction potential between A + BC, B+CA or C+AB fragments is zero. 

We will define the final states of the products as
\begin{eqnarray}\label{eq:final-states}
     \varphi_{\beta,v',j',\Omega'} = \sum_{\Lambda'} W^{Jp}_{\Omega'\Lambda'} (\phi',\theta',\chi') {Y_{j'\Omega'} (\gamma',0)\chi_{v'j'}(r')\over r' R'},
\end{eqnarray}
which are eigenstates of the products at long $R'$ values, 
where it is assumed that the diabatic electronic states are not coupled on the reactants while for products the electronic states may be coupled. Therefore, the rovibrational states of the products are in general calculated in a coupled set of electronic states~\cite{Sanz-Sanz-etal:21}, and the quantum number $v'$ is then used to denote the vibrational and electronic states.

Defining the projection coefficient at Chebyshev iteration $k$ as 
\begin{eqnarray}\label{eq:Chebyshev-products-overlaps}
    C_{v'j'\Omega'\beta}(k) = \int dV_{\beta} \delta(R'-R'_\infty) \left\langle \varphi_{\beta,v',j',\Omega'} \Bigg\vert \Psi^{Jp}(k)\right\rangle,
\end{eqnarray}
the $S$-matrix elements are defined as~\cite{Balint-Kurti-etal:90,Gray-Balint-Kurti:98,Gomez-Carrasco-Roncero:06}
\begin{eqnarray}\label{eq:S-matrix-from-chebyshev}
    S^{Jp}_{v_0j_0\Omega_0, \beta v',j'\Omega'} (E) = -i \sqrt{2K'\over \pi\mu'} {e^{-iK'R'} \over a_0(E)} \sum_k c_k({\hat H}_s,E)\, C_{v'j'\Omega'\beta}(k).
\end{eqnarray}
with $K'(E)=\sqrt{ 2 \mu' (E-E_{v'j'\Lambda'})/\hbar^2}$ with $\mu'$ and $E_{v'j'\Lambda'}$ being the reduced mass and diatomic eigenvalues for products.

The problem to evaluate the coefficients in Eq.~\eqref{eq:Chebyshev-products-overlaps}
is that in general the two wave functions are written in different Jacobi coordinates, one corresponding to reactants and the other to products. 

The use of either reactants or products Jacobi coordinates, requires the use of large grids to describe at the same time the two asymptotic regions.
One alternative, when the reaction is direct and does not present
any wells, is the use of the so called Reactant-Product Decoupling (RPD) method
~\cite{Peng-Zhang:96, Zhu-etal:97, Dai-Zhang:97,Althorpe-etal:97,Althorpe:01}. In this method, the wave packet is split  in reactant
and product channels, and its evolution is
calculated separately in the best suited coordinates.
The flux between different channels is exchanged using Negative Imaginary Potentials~\cite{Dai-Zhang:97,Althorpe:01}. Transforming the coordinates is non-trivial, as a straightforward approach would nearly double the number of coordinate indices required (five in this case: $ r, R, \gamma, \Omega,$ and $\Lambda$). To illustrate the computational cost of this transformation, applying the Hamiltonian to the wave packet involves $5+1$ nested loops.

In a more general case, to get the state-to-state reaction probabilities only one set of Jacobi coordinates is chosen.
In these cases, the internal grids have to be larger to cover all
the rearrangement channels. There are two alternatives:
\begin{enumerate}
    \item Products coordinates based (PCB) method: the initial wave packet, written as a product of monodimensional functions in reactants coordinate, is transformed to products. In this case the overlap required  in Eq.~\eqref{eq:Chebyshev-products-overlaps} can be done directly.

    \item Reactants coordinates based (RCB) method: in which the wave packet 
    at each Chebyshev iteration has to be transformed. To do so, a sequential transformation, passing through intermediate coordinates, is done  as it has been proposed~\cite{Gomez-Carrasco-Roncero:06}.
\end{enumerate}

The choice between PCB and RCB depends on the masses involved in the 
mass combination as discussed in detail in Ref.~\cite{Gomez-Carrasco-Roncero:06}. Thus for heavy-light + heavy' $\rightarrow$ heavy-heavy' + light the RCB  method is very effective. In the case of heavy-light+heavy' $\rightarrow$ heavy +light-heavy' case PCB is better. In the intermediate case, the choice is flexible, with the symmetry of identical atom permutations playing an important role.

{The parameter {\bf iprod} in the {\bf inputprod} namelist controls whether only the total flux is computed ($= 0$), computes the state-to-state probabilities using products Jacobi coordinates ($=1$) or computes the state-to-state probabilities using reactant Jacobi coordinates ($=2$).}

\subsection{Cross sections}

The state-to-state cross sections for reactive and inelastic collisions are obtained in the usual partial wave summation as
\begin{eqnarray}\label{eq:collisional-xsection}
\sigma_{v_0j_0\Lambda_0,\beta v',j'} (E) = {\pi\over K_0(E)^2 (2j_0+1)} {\cal C}_{v_0j_0\Lambda_0,}\beta v',j' (E)
\end{eqnarray}
where the cumulative probability, $ {\cal C}_{v_0j_0\Lambda_0,\beta v',j'} (E)$ is defined as
\begin{eqnarray}\label{eq:cumulative-probability}
    {\cal C}_{v_0j_0\Lambda_0,\beta v',j'}(E) =\sum_J\sum_p \sum_\Omega \sum_{\Omega_0}
(2J+1) \left\vert  S^{Jp}_{v_0j_0\Omega_0, \beta v',j'\Omega'} (E)  \right\vert^2.
\end{eqnarray}
This expression needs to have the individual S-matrix elements 
 for all $J$, $p$ and $\Omega_0$ calculated previously. Following the J-shifting approach~\cite{Bowman:85}, the individual probabilities are shifted up as $J$ increases.

 The cumulative probabilities can be approximated by considering the so-called J-shifting interpolation method~\cite{Gomez-Carrasco-etal:05,Zanchet-etal:10b,Aslan-etal:12}. In this method the reaction probabilities are obtained for some selected total angular momenta, $J_i$, so that for each $\Omega_0$ and $p$ family, the probabilities for $J$ can be obtained as a linear interpolation
 as
 \begin{eqnarray}\label{eq:J-shifting-interpolation}
     \left\vert S^{Jp}_{i,f} (E) \right\vert^2 &=&{J-J_1\over J_2-J_1}  \left\vert S^{J_1p}_{i,f}(E- B\left\lbrack J(J+1)-J_1(J_1+1)\right\rbrack) \right\vert^2  \nonumber \\
     &+&{J_2-J\over J_2-J_1}  \left\vert S^{J_2p}_{i,f} (E + B\left\lbrack J_2(J_2+1)-J(J+1)\right\rbrack) \right\vert^2 
 \end{eqnarray}
 with $J_1\leq J\leq J_2$ and $B$ rotational constant is previously fitted for each $\left\lbrack J_1,J_2\right\rbrack$ interval.

 The cumulative reaction probabilities and cross sections for reactive and inelastic processes
 are calculated with the auxiliary programs {\bf CRP-fast.f} and {\bf CIP-fast.f}. If 
 only total reaction probabilities were calculated, the total cross section (summing over all final states in Eq.~\eqref{eq:collisional-xsection}) are obtained with the program {\bf sigmaFromS2prod.f}.

 The total photodissociation cross section for a $J_i\rightarrow J_f$ rotational transition is readily obtained from the autocorrelation function using Eqs.~\eqref{eq:fourier-transform} and \eqref{eq:chebyshev-transform}
 \begin{eqnarray}\label{eq:spectrum}
 \sigma( h\nu) &=& {A h\nu \over 2\pi\hbar} 2{\cal R}\int_0^\infty dr\,\left\langle   \Psi^{JMp\leftarrow J_ip_i}(t=0) \vert \Psi^{JMp\leftarrow J_ip_i}(t)\right\rangle  
 \\
  &=& {A h\nu \over 2\pi\hbar}  \sum_{k=0} 2{\cal R} c_k({\hat H}_s,E)\,\left\langle   \Psi^{JMp\leftarrow J_ip_i}(k=0) \vert \Psi^{JMp\leftarrow J_ip_i}(k)\right\rangle 
      \nonumber
\end{eqnarray}
with $A=1/\hbar^2 \epsilon_0 c$ and ${\cal R}$ denoting the real part. This is done 
with the {\bf cheby-spectra.f} auxiliary program. 
    
\section{Distribution}\label{sec:distribution}

The code can be downloaded from the git repository:

  \texttt{git clone \repourl}

The repository and the program auxiliary file have the following structure:
~
\dirtree{%
.1 madwave3.
.2 BIN\DTcomment{Holds binary files after compilation. Contains the compilation scripts or Makefile}.
.2 EXAMPLES\DTcomment{Examples for different applications}.
.2 SRC\DTcomment{Source code files}.
.2 .gitignore.
.2 README.md\DTcomment{Minimal usage comments}.
}
~

To compile the code navigate to the {\bf BIN} directory and run either the Makefile or bash script {\bf colmad3.sh} to compile the {\bf mad3.out} code or {\bf bnd3.out} devoted to calculation of bound states. Several other auxiliary codes are compiled in this step, which will be indicated latter. In order to compile all these codes with the Makefile type ``{\texttt make main aux}''.
The compilation requires two libraries: MPI (we use OpenMPI) and FFTW3 (the implementation in MKL does not work with the present set of parameters). The code has been tested with gfortran and Intel Fortran Compiler compilers. In new versions of gfortran, the use of {\bf include ``mpif.h''} is deprecated, and without changing the codes
this problem can be avoided by including the flags in the compilation (using Makefile or colmad3.sh): {\bf FCFLAGS=``-DUSE\_MPIF -fallow-invalid-boz -fallow-argument-mismatch''}.

The FFTW3 library is not thread-safe. For this reason the use of OpenMP is not
direct, and a procedure has been implemented to solve this problem. Therefore, the use of OpenMP should be checked. Once this is done, the options to use OpenMP (for example, ``-fopenmp'' when using gfortran) have to be added to the 
Makefile or colmad3.sh procedures.

In the directory {\bf EXAMPLES} some examples are provided, discussed below, which use some potential and electric dipole moments in directory {\bf PES}.

The two main programs, {\bf mad3.out} and {\bf bndgrid.out}, use some system dependent data that must be prepared in a preparation stage: the potential, in directory {\bf pot} and the matrix dipole moments, in directory {\bf dip}.
In addition,  {the rovibrational states of diatomic fragments for} reactants and products are calculated and kept in directory {\bf func}. The three directories must be created and empty before execution of program {\bf pot.out},
which must be compiled with the Makefile or script {\bf colpot.sh} using system dependent subroutines provided by the user, as described below.

{

\section{System dependent programs provided by the user}\label{sec:dependent_programs}

The user must provide three routines for the {\bf pot.out} program to generate the data required to run MADWAVE3:
{\bf setxbcpotele}: This routine supplies information about the electronic states and initializes the data needed to call the potential in the second routine, {\bf potelebond}. In the case of photodissociation, also the transition dipole moments need to be provided in the {\bf dipele} routine. The arguments of these three routines
are explained below:

\begin{minted}[frame=single]{fortran}
subroutine setxbcpotele(iomdiat,iomatom,sigdiat,sigatom, &
                        nelec,nelecmax)
    ! Initialize variables of the user provided PES.
    ! iomdiat: Projection of the orbital electronic 
    !          angular momentum (Lambda) for each of 
    !          the nelec states considered (or Omega 
    !          in Hund's case c))
    ! iomat: Similar to iomdiat but for the atom.
    ! sigdiat: Parity for the reflection of electronic 
    !          wave function through the xz-body fixed
    !          fram of the diatomic fragment.
    ! sigatom: Similar to sigdiat but for the atom.
    ! nelec: Number of electronic states considered.
    ! nelecmax: Total number of electronic states in the
    !           PES, usually equal to nelec.
    implicit none
    integer, intent(in) :: nelec, nelecmax
    integer, intent(out) :: iomdiat(nelecmax)
                            iomatom(nelecmax)
                            sigdiat(nelecmax)
                            sigatom(nelecmax)
    ! Set values
    ! If required, execute the PES initialization
end subroutine setxbcpotele
\end{minted}

\begin{minted}[frame=single]{fortran}
subroutine potelebond(r10, r02, costet, potmat, nelec, & 
                      nelecmax)
    ! Compute the potential energy matrix.
    ! r10, r02: Diatomic internuclear distances for the 
    !           AB and BC fragments. In Bohr.
    !           When iprod = 1, AB+C are products.
    ! costet: Cosine of the angle formed by r01 and r02.
    ! potmat: Potential energy matrix in the diabatic
    !         representation. In Hartree.
    ! nelec: Number of electronic states considered.
    ! nelecmax: Total number of electronic states in the
    !           PES, usually equal to nelec.
    implicit none
    integer, intent(in) :: nelec, nelecmax
    real(8), intent(in) :: r10, r02, costet
    real(8), intent(out) :: potmat(nelecmax, nelecmax)

    ! Fill potmat with a user defined PES
end subroutine potelebond
\end{minted}

\begin{minted}[frame=single]{fortran}
subroutine dipele(rp, Rg, cgam, dx, dy, dz, nelec, nelecmax)
    ! Compute the electronic transition dipole moments.
    ! rp, Rg: Distances in Jacobi coordinates. 
    !         Small and big rs.
    !         In Bohr.
    ! cgam: Cosine of the angle between r and R.
    ! dx, dy, dz: Electronic transition moments from a single
    !             initial electronic state to the nelec 
    !             electronic states considered, for the 
    !             x, y, z components of the Jacobi 
    !             body-fixed frame.
    !             In atomic units, NOT Debye.
    ! nelec: Number of electronic states considered.
    ! nelecmax: Total number of electronic states in the
    !           PES, usually equal to nelec.
    implicit none
    integer, intent(in) :: nelec, nelecmax
    real(8), intent(in) :: rp, Rg, cgam
    real(8), intent(out) :: dx(nelec), dy(nelec), dz(nelec)

    ! Fill dx, dy, dz
end subroutine dipele
\end{minted}

To execute the program {\bf pot.out}, the directories {\bf pot}, {\bf func} and {\bf dip} (if needed) must be in the parent directory with respect to the working directory ($i.e.$ as {\bf ../pot}, etc)
}

In {\bf potelebond} the input variables r10, r02, costet determine the geometry
of the triatomic system. Often, the potentials developed for triatomic systems
are expressed as:
\begin{enumerate}
    \item Internuclear distances r10, r12, r02: in this case the missing r12 distance is directly obtained as

    \begin{equation}
        r12= \sqrt{r10^2+r02^2 - 2 r10\, r20\, \cos{\gamma}}
    \end{equation}
    where $\cos{\gamma}$ is in the costet variable.

    \item In Cartesian coordinates for the three atoms $x_0, y_0, z_0, x_1, y_1,z_1,x_2, y_2, z_2$: Assuming the three atoms in the x-z plane, with atom 0 at the origin 
    and atom 1 along the z-axis, the Cartesian coordinates are then obtained as
    \begin{eqnarray}
        z_1=r10, \quad z_2= r02\, \cos(\gamma) \quad x_2= r02\,\sin(\gamma),
    \end{eqnarray}
    after initializing all Cartesian coordinates to zero.
\end{enumerate}
Many triatomic systems have been studied, with many different mass combinations, in collisions such as Li+HF$\rightarrow$ LiF+H \cite{Aguado-etal:97b,Krasilnikov-etal:13}, O+HF$\rightarrow$ F+OH \cite{Gomez-Carrasco-etal:05,Zanchet-etal:10b}, C$^+$+H$_2$\cite{Zanchet-etal:13}, S$^+$ +H$_2$\cite{Zanchet-etal:19,Zanchet-etal:24}, Ca+HCl\cite{Sanz-etal:05}, H+H$_2^+$\cite{Sanz-Sanz-etal:10}, H$^+$+H$_2$ \cite{Gonzalez-Lezana-etal:05}, F+DCl\cite{Bulut-etal:15}, 
N+OH \cite{Bulut-etal:11}, H$^+$ +LiH \cite{Aslan-etal:12}, and photoinitiated processes in LiHF \cite{Paniagua-etal:99,Aguado-etal:03}, ArI$_2$\cite{Roncero-etal:01},  HCN\cite{Chenel-etal:16,Aguado-etal:17}, OHF$^-$\cite{Gonzalez-Sanchez-etal:04,Gomez-Carrasco-etal:06}, etc.

The program {\bf pot.out} calculates the bound states of the reactant and products diatomic fragments. The diatomic potentials of the reactants (01) and products (02) are printed in the files {\bf potr.eEE} and {\bf potdia-prod02.dat}, where (EE denotes the electronic state).
The examination of these potentials, together with the eigenvalues (in salpot.000) and eigenfunctions (in wv01-ielecEE-j00.dat and wv02-vVVj0.dat) of 01 and 02 diatomic fragments provides a simple way to check the implementation of a new PES. Also, for J=0 and setting iwrt\_pot=1 (in the namelist INPUTWRITE described below), the execution of program {\bf mad3.out} prints the PES in the files {\bf potr1r2.eEE.FF}, in the grid of
points defined in inputwrite, as described below. Note, that this potential is shifted
so that the energy of the reactant level determined by IELECREF,NVREF,JREF,IOMREF 
has zero energy.

In the case of several coupled electronic states, a diabatic representation is assumed\cite{Smith:69,Baer:80,Mead-Truhlar:82,Koppel-etal:84,Thiel-Koppel:99,Koppel-cap4-conicalbook:04,Gomez-Carrasco-etal:06,Guo-Yarkoni:16,Naskar-etal:20}. The diabatic represention is built by a unitary transformation from the usual adiabatic representation. It allows the inclusion of the coupling among different electronic states through a slowly varying potential term. The singularity of non-adiabatic couplings
at conical intersections is avoided in the diabatic representation, thus presenting computational advantages. Diabatic representations are not unique and cannot remove all the non-adiabatic terms in polyatomic systems~\cite{Mead-Truhlar:82}, but they allow to regularize the singularities at conical intersections~\cite{Thiel-Koppel:99}. There is no standard method to generate the diabatic states, specially beyond two electronic states and there is a pleiad of methods \cite{Smith:69,Baer:80,Mead-Truhlar:82,Koppel-etal:84,Thiel-Koppel:99,Koppel-cap4-conicalbook:04,Gomez-Carrasco-etal:06,Guo-Yarkoni:16,Naskar-etal:20}.

In general, it is assumed that the diabatic states correlate to the adiabatic ones at long distances. In such situation, there is no coupling among different electronic states, and the rovibrational states of the diatomic fragments in each rearrangement could be calculated independently in each electronic state. When dealing with reactions, this is not always the case. For example, the diabatic states of H$_3^+$ 
can be formed in a simple basis in which each electronic state corresponds to a charge in a single H atom \cite{Aguado-etal:21}. This is true when studying H$_2$ + H$^+$. However, for H$_2^+$ + H this is not the case, and a choice has to be made in which rearrangement channel the diabatic and adiabatic representations coincide at long distances, as it was done recently in the study of non-adiabatic transition in H$_2^+$ + H reactive collisions \cite{Sanz-Sanz-etal:21}. For this reason, in MADWAVE3 the rovibrational states for the 02 products are calculated in the full manifold of electronic states, and in  wv02-vVVj0.dat files each vibrational state VV is represented as a linear superposition in the different electronic states.

\section{Input data}\label{sec:input_data}

The rest of the input data is provided in different namelists in the file {\bf input.dat} that must be in the working directory. Below the parameters
of the different namelists are listed and briefly explained in Table~\ref{tab:parameters}.

\begin{longtblr}[
  caption = {Brief description of the data included in each namelist in {\bf input.dat}.
    Distances in \AA, energies in eV. 
    Variables with initial letter starting k-q are integer, otherwise they are real*8.
    Default values are given in parenthesis},
  label = {tab:parameters},
]{
  colspec = {|XX[1.3]|},
  row{even} = {gray9},
} 
\hline
{\bf inputgridbase}  & \\\hline
npun1,rmis1,rfin1     & $r$ no. of points in the interval \lbrack rmis1, rfin1\rbrack \\
npun2,rmis2,rfin2     &  $R$ no. of points in the interval \lbrack rmis2, rfin2\rbrack \\
nangu & no. of Gauss Legendre quadrature points for $\gamma$ \\
Jtot & total angular momentum, $J$\\
iparity & total inversion parity, $p$\\
jini,jmax & first and last reactants BC angular momentum for flux evaluation \\
inc & =1, 2 for hetero or homo nuclear BC molecules\\
j0 & initial value of j when inc=2 \\
nelecmax & no. of electronic states \\
iomini(0) , iommax & $\Omega$ values in the expansion of wvp, Eq.~(\ref{eq:total-function})\\ 
nvini,nvmax & first and last vibrational levels of BC reactants \\
nvref,jref,iomref,ielecref & $v_0,j_0,\Omega_0,\Lambda_0$ reactants quantum numbers, Eq.~(\ref{eq:initial-collision-wvp}). Its energy is taken as the zero of energy.\\\hline

{\bf inputprod} & \\ \hline
iprod (0) &  {= 0 only total flux \\
        = 1 products coordinates for state-to-state prob.
        for levels defined by jini,jmax,nvini,nvmax  \\
        = 2 using reactant coordinates for state-to-state prob. products levels defined below}\\
nviniprod,nvmaxprod & first and last vibrational states of products, $v'$ in Eq.~\eqref{eq:final-states}\\
jiniprod,jmaxprod & first and last rotational states of products, $j'$  in Eq.~\eqref{eq:final-states}\\
iomminprod(0),iommaxprod(0) & initial and final $\Omega'$ of products \\
Rbalinprod & $R'_\infty$ at which the flux on products is done, Eq.~\eqref{eq:Chebyshev-products-overlaps}\\
n2prod0(1),n2prod1(npun2) & initial and final grid points to represent. 
                          Vib. functions in intermediate representation 
                          (to save memory and reduce coordinate transformation time).\\

                          \hline
    {\bf inputprocess} & \\\hline
    iphoto & = 0 collision, =1 photodissociation \\
    \hline
    
    {\bf inputtime} & \\\hline
    ntimes & number of Chebyshev iterations, $k$, per loop, Eq.~(\ref{eq:wvp-k}) \\
    nloop & number of loops \\
    kminelastic & initial $k$ at which the elastic flux is evaluated \\
    \hline
    {\bf inputflux} & \\
    \hline
    r1flux\_ang & $r^*$ value at which flux is evaluated, (=absr1-1) Eq.~(\ref{eq:total-flux}) \\
    netot & number of energies at which flux is evaluated \\
    ekinmin\_eV & minimum energy for flux evaluation \\
    ekinmax\_eV & maximum energy for flux evaluation \\
    ncontfile & {=1 for long propagations to restart calculations. \\
        Restarting files are written at the end of each loop}\\
\hline
    {\bf inputwrite} & \\ \hline 
    iwrt\_pot(0) & 
    {if 1 writes the potential in files potr1r2.eX.Y for $r_k$, $R_m$, $\gamma_n$ \\
                    $k$=1,1+n1plot,1+ 2 n1plot, ..., npun1 \\
                     $m$= 1,1+n2plot, 1+2 n2plot,..., npun2 \\
                     $n$=1,1+nangplot, 1+2 nangplot, ...,nangu} \\
    iwrt\_wvp & if =1 write the wave packet for $r_k$, $R_m$, $\gamma_n$ at each loop \\
    iwrt\_reac\_distri & {= 0 don't write reactants distribution (except iprod=1) \\
                       = 1 writes reactants probabilities in files \\
                         the file is overwritten at each loop \\
                       = 2 writes overlap of Eq.~(\ref{eq:Chebyshev-products-overlaps}) \\
                         to re-process the final distribution (needed when icontfile=1)} \\ 
    n1plot,n2plot,nangplot & defined for iwrt\_pot \\
    \hline
    {\bf inputpotmass} & \\
    \hline
    system & character*20 ``name of system'' \\
    xm1,xm0,xm2 & masses in amu of atoms 01 + 2 $\rightarrow$ 1 +02 \\
    VcutmaxeV & maximum values in eV of potential energy \\
    radcutmaxeV & maximum values in eV of radial kinetic energy \\
    rotcutmaxeV & maximum values in eV of rotational energy \\
    R2inf\_radial\_functions & distance $R$ to calculate reactant vibrational states (100 bohr)\\
     R1inf\_radial\_functions & distance $R$ to calculate products vibrational states (100 bohr)\\
   \hline
    {\bf inputbnd} & \\
    \hline
    Jtotini,iparini & total angular momentum $J_i$ and parity $p_i$ of inital bound state \\
    nvbound & vibrational state to read, as generated by {\bf bndgrid.out} \\
    nprocbnd & processors(files) used to calculate initial bound state \\
    maxbnddim& maximum size to the vectors used for bound state \\
    igridbnd & must be 1 if using {\bf bndgrid.out}\\
    \hline
    {\bf inputcol} & \\
    \hline
    Rcolini\_ang & $R_0$ center of initial Gaussian (in \AA) for collisions, Eq.~(\ref{eq:initial-gaussian}) \\
    ecol\_eV & mean collision energy (eV) of initial Gaussian \\
    deltaE\_eV & energy width (in eV) of initial Gaussian. \\
    \hline
    {\bf inputabsorcion} & \\\hline
    absr1,absalp1,n1expo(2) & $r_I$ (\AA), $A_r$ and $n_i$ for $r$ in Eq.~(\ref{eq:imaginary-potential}) \\
    absr2,absalp2,n2expo(2) & $R_I$ (\AA), $A_R$ and $n_i$ for $R$ in Eq.~(\ref{eq:imaginary-potential}) \\
    \hline
\end{longtblr}

\section{Examples and Output\label{sec:example}}

The prototype state-to-state reactive collision 
\begin{eqnarray}\label{Eq:s2s-reaction}
    \text{H} + \text{DH}(v=0,j=0) &\rightarrow& \text{H}_2(v',j') + \text{D} \nonumber
\end{eqnarray}
is studied here, also considering the inelastic process, 
\begin{eqnarray}\label{Eq:s2s-inelastic}
    \text{H} + \text{DH}(v=0,j=0) &\rightarrow& \text{H} + \text{DH}(v',j') \nonumber
\end{eqnarray}
For the exchange process only the total reaction probability and total cross section is calculated:
\begin{eqnarray}\label{Eq:total-exchange}
    \text{H} + \text{DH}(v=0,j=0) &\rightarrow& \text{HD} + \text{H}. \nonumber
\end{eqnarray}

In the following we present a shorter version of the example which can be found in the EXAMPLES folder of the MADWAVE3 repository.

\subsection{Preparatory step}
The first step is to compute the PES and reactants (and products) eigenfunctions. For that, we must specify information about the grid used to represent the PES, the masses the atoms and whether the states of the products should be computed. Here we explain some of the parameters that should be set in the {\bf input.dat} file.

%

\begin{minted}[frame=single]{text}
&INPUTGRIDBASE
! 256 grid points from 0.1 to 20 A 
 NPUN1= 256, RMIS1= 0.1d0, RFIN1= 20.d0, NPUN1MIN= 64, 
 
! 256 points from 1e-3 to 20 A            
 NPUN2= 256, RMIS2= 0.001d0, RFIN2= 20.d0, 
 NANGU= 140 ! 140 angular points
 NELECMAX= 1, ! 1 electronic state
 JTOT= 0, IPARITY= 1
 IOMMIN= 0, IOMMAX= 0,
 J0= 0, INC= 1
 
 ! j=0 to j=39 for reactant's rotational states
 JINI= 0, JMAX= 39, 
 
 ! v=0 to v=5 for reactant's vibrational states
 NVINI= 0, NVMAX= 5, 
 
 NVREF= 0, JREF= 0, IOMREF= 0, IELECREF=  1,
 /
\end{minted}

\begin{minted}[frame=single]{text}
 &inputpotmass
!-----------> masses(in amu) and potential(in eV) cut data
system='H+DH'
xm0=2.014101779d0 ! mass of atom 0
xm1=1.007825035d0 ! mass of atom 1
xm2=1.007825035d0 ! mass of atom 2
VcutmaxeV=2.5d0 ! maximum potential energy
radcutmaxeV=2.5d0
rotcutmaxeV=5.d0
/
\end{minted}

\begin{minted}[frame=single]{text}
&inputprod
iprod=2  ! =0 no products state distribution,
         ! =1 using products Jacobi coordinates,
         ! =2 using reactants Jacobi coordinates,
         !transforming to products to extract s2s properties
         
! v=0 to v=5 for product's vibrational states
nviniprod=0,nvmaxprod=5

! j=0 to j=20 for product's rotational states
jiniprod=0, jmaxprod=20

iomminprod=0!  , iommaxprod=0 ! Jtot
Rbalinprod=6.d0
n2prod1=256 !npun2/2
/
\end{minted}

In this example we employ an in-house PES provided in the MADWAVE3 repository which reproduces
very well the PES by Pendergast (BKMP2). This potential presents a barrier to 
reactions of $\approx$ 0.25 eV, and the study
will be performed up to 1.5 eV above the H + DH$(v=0,j=0)$ initial state,
which determines the zero of energy. Final states up to v=5 and j=40 states
will be considered for reactants and products. Also, potential will be cut
to {\bf vcutmax\_eV}= 2.5 eV.

{
Our H$_3$ PES is invoked through the \texttt{VH3} subroutine, which receives as input the three interatomic distances.
\begin{minted}[frame=single]{fortran}
subroutine VH3(RAB,RBC,RAC,VTOT,DVTOT,ID)
    ! RAB, RBC, RAC: interatomic distances
    ! VTOT: Electronic energy
    ! DVTOT: Derivatives of the energy w.r.t. distances
    !        (not implemented)
    ! ID: status of computation
    use H3pot ! module with actual PES.
    implicit none
    real(8), intent(in) :: RAB, RBC, RAC
    real(8), intent(out) :: VTOT
    real(8), dimension(3), intent(out) :: DVTOT 
    integer, intent(out) :: ID
    ! Computation...
end subroutine VH3
\end{minted}

In this particular case, the routines which interface the PES with MADWAVE3 (\texttt{setxbcpotele} and \texttt{potelebond}) are defined in the \texttt{pothead.f} file:

\begin{minted}[frame=single]{fortran}
 subroutine setxbcpotele(iomdiat,iomatom,sigdiat,sigatom
&                       ,nelec,nelecmax) 
   implicit real*8(a-h,o-z)
   dimension iomdiat(nelecmax),iomatom(nelecmax)
   dimension sigdiat(nelecmax),sigatom(nelecmax)

   write(6,*)'Potential setting message'

   ! Check that only one electronic state is being used.
   if(nelecmax.ne.1)then
        write(6,
&   '("  This PES is prepared for a single state ")')
        stop
   endif
   nelec=1
   iomdiat(1)=0
   iomatom(1)=0
   sigdiat(1)=+1
   sigatom(1)=+1
   return
  end setxbcpotele

 subroutine potelebond(r1,r2,costet,potmat,nelec,nelecmax)
   use H3_pot
   implicit real*8(a-h,o-z)
   dimension potmat(nelecmax,nelecmax)
   dimension der(3)
   rref=1.4d0
   rasi=100.d0
   call  VH3(rasi,rasi,rref,pot,der)
   vref=pot

   ! Compute the third distance.
   r3=r1*r1+r2*r2-2.d0*r1*r2*costet
   r3=dsqrt(r3)
   if(r1.lt.0.25d0)r1=0.25d0
   if(r2.lt.0.25d0)r2=0.25d0
   if(r3.lt.0.25d0)r3=0.25d0
   call VH3(r3,r2,r1,pot,der)
   potmat(1,1)=pot-vref
   return
  end potelebond
\end{minted}
The {\bf pot.out} code can be compiled with the Makefile provided in the package.
}

The empty directories {\bf pot} and {\bf func} are created prior to the execution of {\bf pot.out} code, which will compute the PES in the selected grid and the reactant's and product's wavefunctions.

\subsection{Reaction probabilities at $J=0$}

Usually, the calculations for $J=0$ are first converged prior to calculating
the remaining $J$'s. This requires to check the parameters of the grids until
full convergence of the different energy-resolve state-to-state probabilities
is reached. This of course implies to check how these quantities converge
with the number of Chebyshev iterations. The parameters of the calculations
are printed in files {\bf sal.XYZ}, where XYZ runs from 0 to the number of 
processors minus one. Once the propagation starts, the norm of the Chebyshev
wave packet at iteration $k$, $\Psi(k)$, is printed only in sal.000 file. 
These norms start oscillating, 
until they reach a plateau with a norm = 1/2. Due to the absorption required at
the edges of the grids, this norm decreases progressively until zero, giving information
on how absorption proceeds and how much remains in the physically relevant region. It should be noted, that only if the propagation is very bad it may present
rippling (very bad absorption) or a fast increase (probably because the maximum
value energy is not properly set and the scaled Hamiltonian is wrong).

In order to keep the Chebyshev components of the wave packet real, the initial
one is built as a linear superposition of incoming and outgoing waves. The outgoing
half rapidly evolves towards the edges, resulting in a decrease of the norm from  1/2
to 1/4. The incoming wave packet takes a larger number of iterations to reach 
the edges, since it has to evolve towards shorter distances and then bounce back.

Outgoing and incoming wave packets may interfere in the initial channel of
reactants. To minimize it the parameter {\bf kminelastic} in the elastic probability has to be set to a value where the norm of the component is $\approx$ 1/4.

In the typical workflow we create a new folder for each total angular momentum $J$ and copy the {\bf input.dat} file there. We should have the following directory structure:
~
\dirtree{%
.1 project/.
.2 input.dat.
.2 pot/.
.2 func/.
.2 J000/.
.3 input.dat.
}
~

\begin{figure}[hbtp]
    \centering
    \includegraphics[width=0.8\linewidth]{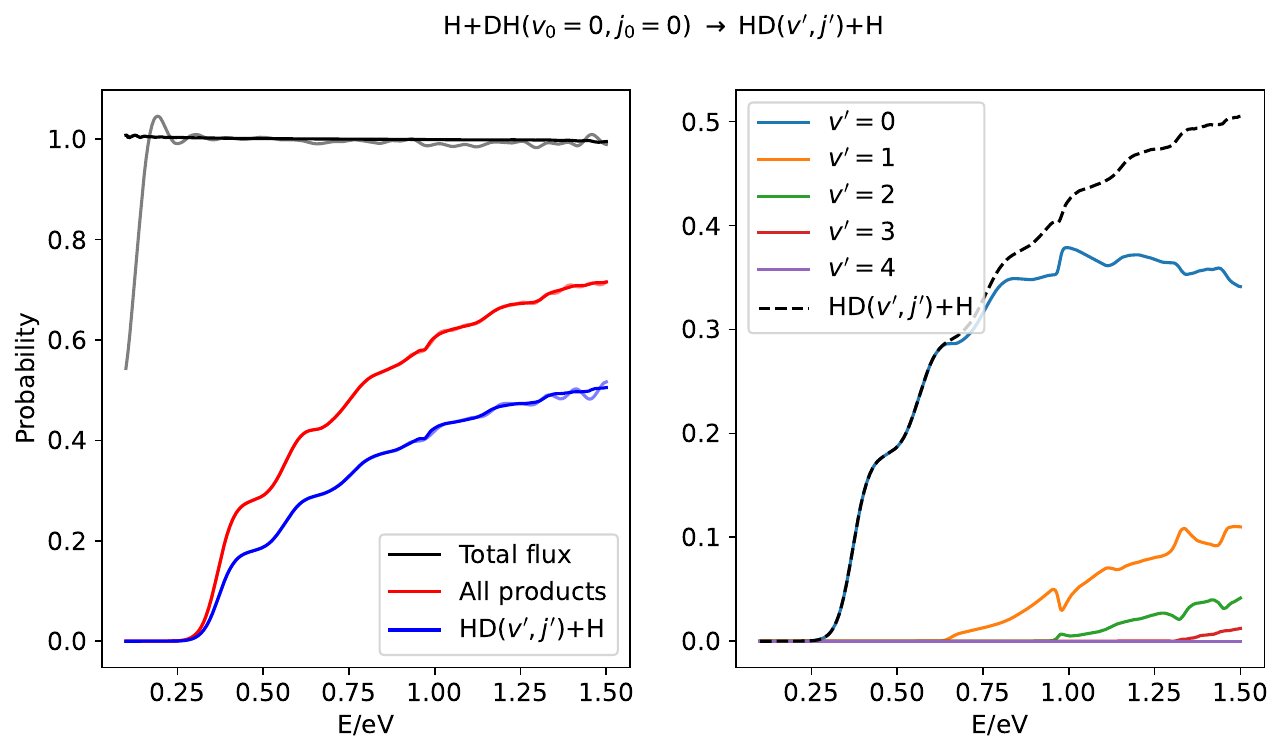}
    \caption{\label{fig-total-probJ0}Left panel presents the total flux in black (column 3 of files S2prod.v00.J000.k\{i\}, total reaction probability (column 2) in red and reaction probability towards the H exchange (column 4) in blue for two different Chebyshev iterations with respect to the energy (column 1). The transparent lines correspond to the second iteration (i=00002) while solid lines correspond to the eight iteration, where convergence has been reached. The right panel breaks down the individual contributions to the total exchange probability in terms of the products vibrational state (columns 5 to 9, for v' = 0  to 4).}
\end{figure}

Next, we navigate to the {\bf J000} folder and run the MADWAVE3 program as {\bf mpirun -n \textless number of MPI tasks \textgreater /path/to/madwave3/mad3.out}. The following files are written after execution:
\begin{itemize}
    \item sal.XYZ: Main output file. Writes the norm of the Chebyshev components at each iteration.
    \item gaussE: Initial wavepacket in the energy domain. First column is the energy and second the value of the wavefunction.
    \item gaussR: Initial wavepacket in the position domain. First column is the $R$ coordinate and second the value of the wavefunction.
    \item distriS2reac.elec\{ielec\}: Total inelastic probability. Represents energy vs probability.
    \item distriS2reac.v\{$v'$\}.e1: Inelastic probabilities for reactants in $v'$. The first column is the energy and the subsequent columns are the probabilities for increasing values of $j'=0, 1, \ldots$
    \item S2prod.v\{$v_0$\}.J\{$j_0$\}.k\{i\}: Reaction probabilities in the $ith$ iteration. To see the last results take the largest value of $i$. First column is the energy, second column the total reaction probability for all the possible product channels. Third is the total probability for reactant and products channel (it should be unity). The fourth is the total reaction probability for the reaction $01+2 \rightarrow 02 + 1$ when {\bf iprod=2}. The subsequent columns present the reaction probability for each of the final vibrational product states for the reaction $01+2 \rightarrow 02 + 1$, summed over all the rotational states. Plotting these files provide a good check of the convergence as shown in Fig.~\ref{fig-total-probJ0}.
    \item potr1r2.e\{ielec\}.\{jelec\}: Contains the ielec, jelec matrix element of the PES on the grid. First column is $r$ and the second is $R$. Subsequent columns are the values of the PES for the various angles in the grid.
    \item Cvjprod.\{i\}: Cumulative reaction probability~\eqref{eq:cumulative-probability}.
    \item prodwv.v\{$v_0$\}.Omg\{$\Omega_0$\} (if iwrt\_pot=1): products rovibrational wavefunction in the vibrational state $v_0$ represented in a bidimensional grid ($r,\gamma$) in reactants Jacobi coordinates.
\end{itemize}
%


To resolve the reaction probability for each of the products rotational states the information in the Cvjprod.\{i\} must be reprocessed with the {\bf distri.out} code, producing the {\bf distriS2prod.v\{$v'$\}.e1} files with equivalent information to that found in distriS2reac.v\{$v'$\}.e1. 

\begin{figure}[hbtp]
\begin{center}
  \includegraphics[width=0.6\linewidth]{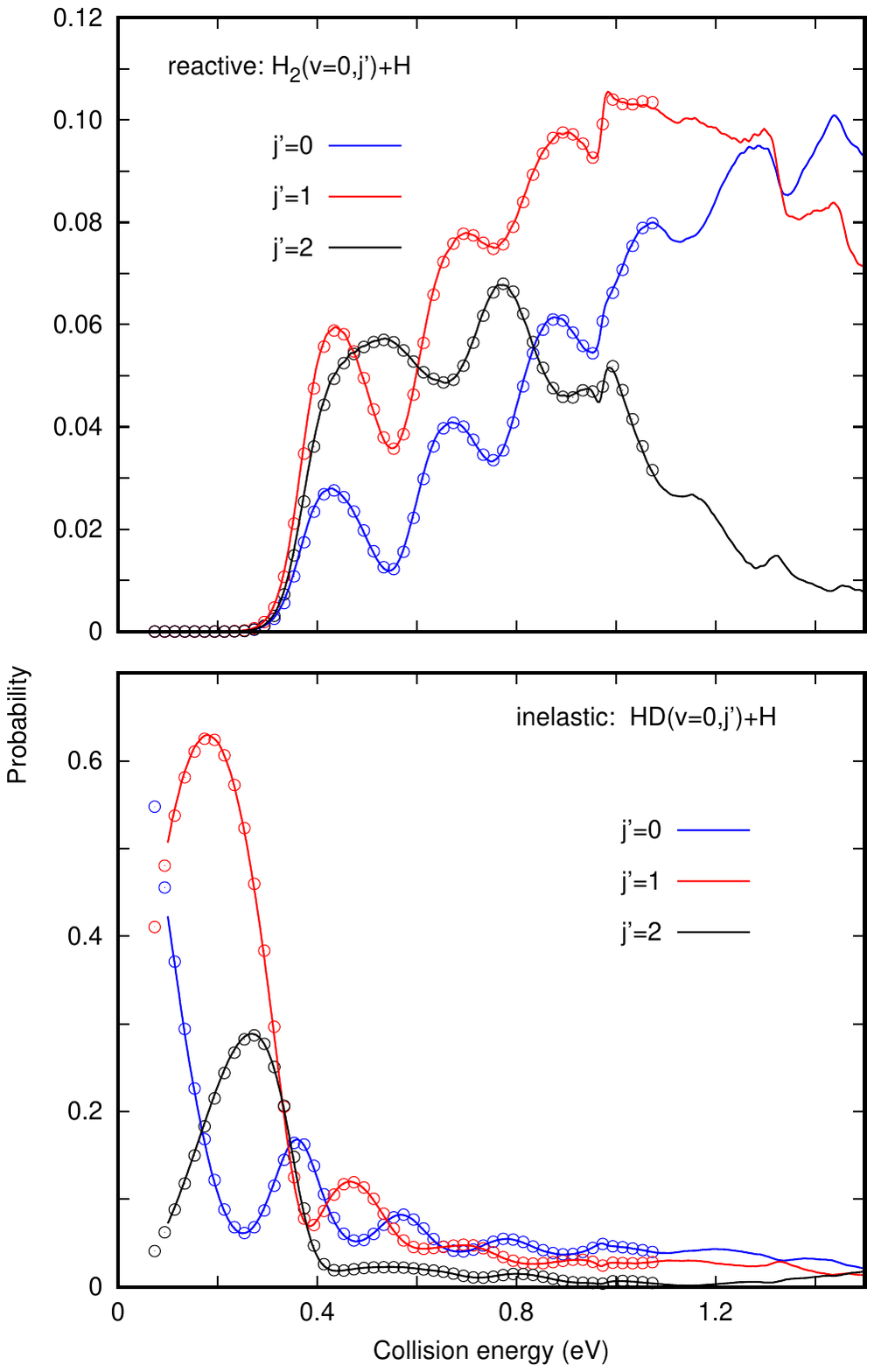}
  
  \caption{{State-to-state probabilities for
  the HD($v=0,j=0$) +H $\rightarrow$ HD($v'= 0, j'=0, 1, 2$) + H inelastic (bottom panel) and
  HD($v=0,j=0$) +H $\rightarrow$ H$_2$($v'= 0, j'= 0, 1, 2$) + D reactive (top panel) collisions for $J$=0. Points are the results obtained with the ABC code, lines are MADWAVE3 results.
   }}    
\label{fig-s2s-probJ0}
\end{center}
\end{figure}

To further check the convergence of individual state-to-state magnitudes, in Fig.~\ref{fig-s2s-probJ0}
the inelastic (top panel) and reactive (bottom panel) reaction probabilities
are compared with hyperspherical time-independent results obtained with the ABC code \cite{Skouteris-etal:00}, shown as points. The agreement between the MADWAVE3 and ABC results is excellent, what demonstrates the convergence and accuracy achieved
with the parameters used (provided in the supplementary information). 

\subsection{Higher partial wave, $J>0$, and cross sections}

\begin{figure}[hbtp]
\begin{center}
  \includegraphics[width=0.6\linewidth]{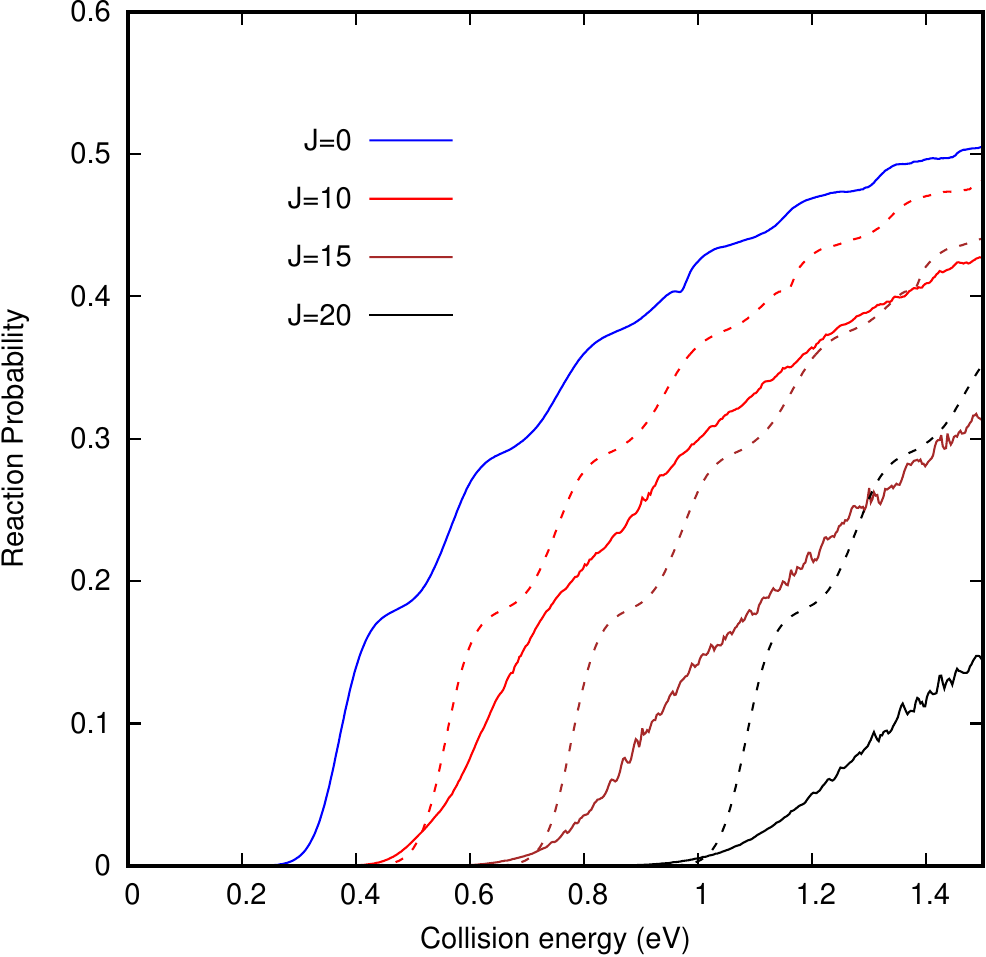}
  
  \caption{{Total H$_2$ reaction probabilities for several $J$. Solid lines
  are MADWAVE3 results, while dashed lines correspond to J-shifting results, $i.e.$,
  obtained from that of $J=0$ as $P^J(E)= P^{J=0}(E+ J(J+1))$ \cite{Bowman:85}, where
  $B$= 1.7 meV is an effective rotational constant.
   }}    
\label{fig:J-shifting}
\end{center}
\end{figure}

To calculate the other partial waves, $J$, the number of helicity components
has to be determined, $\Omega=$ IOMMIN,...,IOMMAX. In this case, convergence is
achieved with IOMMIN=0 and IOMMAX=min($J$,7). As $J$ increases, the effective rotational
barrier increases as well, pushing {the reaction probability} up towards higher energies as shown in Fig.~\ref{fig:J-shifting}. This shifting is qualitatively well explained by the so called J-shifting approach \cite{Bowman:85}. However, the shifted results typically overestimate the exact one, producing too large cross sections. An alternative is the so called J-shifting interpolation, 
in Eq.\eqref{eq:J-shifting-interpolation}.

The cumulative reaction probabilities, Eq.\eqref{eq:cumulative-probability},
are calculated with the {\bf crp.out} program, and reads the namelist {\bf sigmaS2prod} in the input.dat, in addition to those previously
used for the propagations. In this namelist, Jtotmax
and nCalc are the maximum $J$ to be read and the total number of calculated
$J$'s (if all are calculate nCalc=Jtotmax+1). An additional file, called 
{\bf CalculatedJ.dat} is read, with the nCalc values of $J$ (first column) and an effective rotational constant $B$ in eV (second column), 
that optimize the J-shifting interpolation between $J_{calc}$ and $J_{calc}+1$. These files can be found in the CRP directory in the example.

\begin{figure}[hbtp]
\begin{center}
  \includegraphics[width=0.80\linewidth]{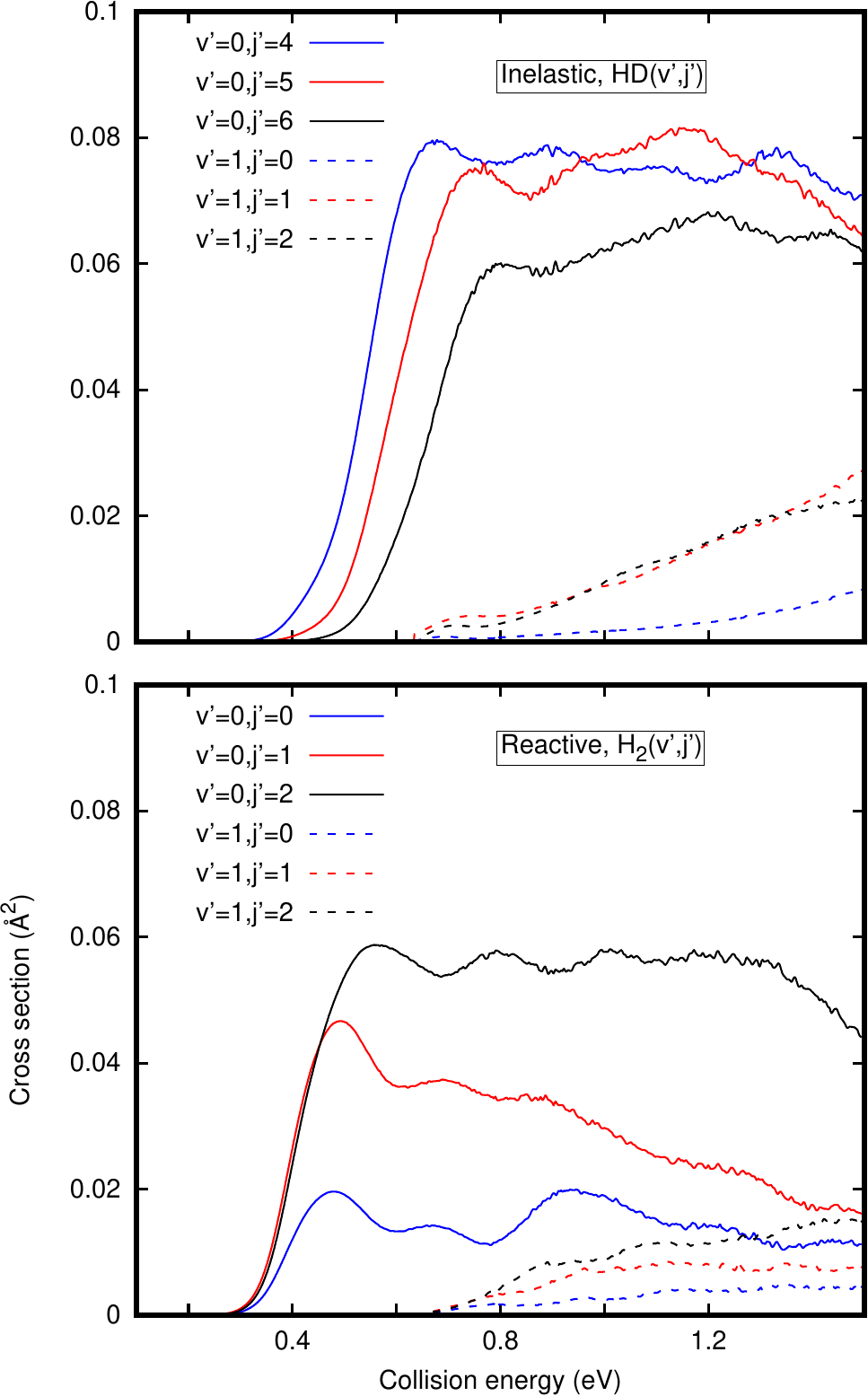}
  
  \caption{{State-to-state cross section for reactive (bottom panel) and inelastic (top panel) for some selected final states of H$_2(v',j')$ and HD($v', j'$) respectively.
  Solid lines are for final $v'$ = 0, dashed lines for $v'$ = 1.
   }}    
\label{fig:s2s-cross}
\end{center}
\end{figure}

We need to create a new directory \texttt{project/}OmgX, where X$=0,\cdots,j_{ref}$, and copy there the reaction probabilities previously calculated into the {\bf JXYZ} files. For $\Omega>$ 0, and extra directory $p$ or $m$ have to be added for +1 or -1 parities, respectively. {After the execution of the {\bf crp.out} program in the CRP directory,} 
the state-to-state reactive 
cross sections are written in files {\bf CRP.vfXX.efY}, where XX refers
to the final vibrational state of products and Y to the final electronic state. { The total cross sections are saved into the {\bf CRPtot.res} files}. A similar description for inelastic cross section is valid but
in {\bf CIP.vfXX.efY} obtained with the {\bf cip.out} program.
In these files, the first column is the collision energy (in eV), the second
is $K_0^2(E)$ (in \AA$^2$), and the remaining are  ${\cal C}_{v_0j_0\Lambda_0,\beta v',j'}(E)$ from $j'=0, 1, 2, ...$. Some state-to-state integral cross section are presented in Fig.~\ref{fig:s2s-cross} for reactive and inelastic case.

\section{Parallelization efficiency}\label{sec:parallelization}

The most efficient parallelization of this code involves distributing each $\Omega$ component of the wave packet across individual processors \cite{Roncero-etal:97}. Since each $\Omega$ is coupled only to adjacent components, $\Omega' = \Omega \pm 1$, this approach limits information transfer to neighboring processors, thereby minimizing communication overhead and maximizing parallelization efficiency. 

To accelerate calculations for low to intermediate $J$ values, the grid in the $\gamma$ angle is also distributed. Due to the L-shape of the grid, not all $\gamma_i$ values share identical grids in $r$ and $R$. To balance memory usage and CPU time, the angular grids are distributed non-uniformly.

Defining $n_\Omega$, $n_\gamma$ and $n_{proc}$ as the number of $\Omega$, the number of angular grid points (nangu in the input) and the number of processors, there are  the following restrictions:
\begin{eqnarray}
    \left\lbrace \begin{array}{ccc}
       if&  n_{proc} \le n_\Omega & \mod(n_{\Omega},n_{proc})=0 \\
        if & n_{proc}> n_\Omega & \left\lbrace\begin{array}{c}
                               n_{proc} = n_\Omega n_{proc}^\gamma\\
                                \mod(n_{proc}^\gamma,n_\gamma)=0
                       \end{array}\right.   
        \end{array}\right.
\end{eqnarray}
meaning that if the number of available processors, $n_{\text{proc}}$, is less than the number of $\Omega$ values, then $n_{\text{proc}}$ must be a multiple of $n_{\Omega}$. If the number of processors exceeds the number of $\Omega$ values, parallelization will also extend to the $\gamma$ angles. In this case, the calculation for each $\gamma$ angle will be distributed across $n_{\text{proc}}^{\gamma}$ processors, where $n_{\text{proc}}^{\gamma}$  must be multiple of $n_{\gamma}$.

The use of MPI libraries allows the use of much larger number of processors than OpenMP. Once an optimal value of processor is attributed to MPI, we can proceed to use the shared memory OpenMP parallelization, with a number of threads typically about 2 to 4. For this reason, below we restrict the discussion to parallelization using MPI in different processors (or cores).

The efficiency of the parallelization depends on the size of the packages that need to be transferred among processors. Therefore the efficiency is very dependent on the system, grids and PESs treated. Below we consider the efficiency for the test study presented in section~\ref{sec:example}. In the output file sal.000 the wall time  is printed in each loop, consisting on ntimes Chebyshev iterations.

\begin{figure}[hbtp]
\begin{center}
  \includegraphics[width=0.60\linewidth]{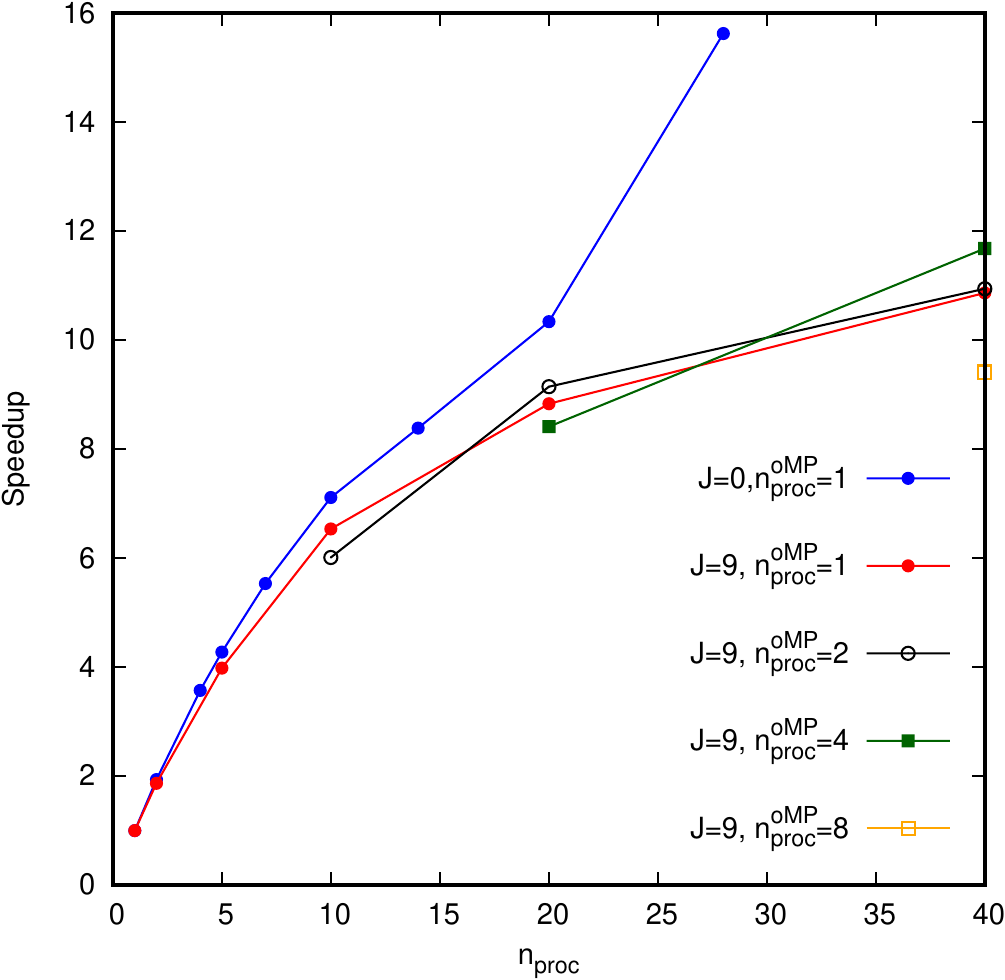}
  
  \caption{{Parallel speedup obtained for J=0 ($n_\Omega=1$)
 and J=9  ($n_\Omega=1$), for different number of processors, $n_{proc}$ and    , $n^{OMP}_{proc}$}}    
\label{fig:efficiency}
\end{center}
\end{figure}

The parallel speedup is defined  as 
\begin{eqnarray}
  S(n_{proc})= { {\rm wallclock\, for\, serial } \over {\rm wallclock\, for\, parallel}},
\end{eqnarray}
and we consider two cases: 
\begin{enumerate}
    \item  $J=0$ only parallelization over the $n_\gamma=$ nangu = 140 angular grid points.

    \item $J=9$, with $\Omega=0,1,...,10$, i.e., $n_\Omega$=10, where a mixed parallelization over $\Omega$ and $\gamma$ is performed
\end{enumerate}
 The parallel speedup is shown in Fig.~\ref{fig:efficiency}. The calculations were performed in a AMD EPYC GENOA (at 4.05 GHz and 1536 GB of RAM), with 2 nodes of 24 cores each, and the compilation was done with gfortran. 

The speedup for $J=$ 0 (blue line) increases continuously up to $n_{proc}$ 20, presenting an extra-speedup from 20 to 28, just when two nodes are used. In this case, only MPI parallelization is used.
For $J=$ 10, the calculations with $n_{\Omega}=10$ are considerably longer in time, passing from 224 s for $J=0$ to 3256 s for $J=10$.
Using only MPI, the calculations for $J$=10 presents a lower speedup. 
This is surprising because the parallelization in $\Omega$ is more efficient (the communication is only among sequential cores), and it is probably due to the increase of size of the communication packets.

The efficiency of combining MPI and OpenMP parallelization (keeping constant the total number of cores, $n_{proc}$, is only shown for $J$= 9, using 1,2 4 and OpenMP threads. For $n_{proc}$ =20 there is a slight increase when using 2 threads, which decreases when using 4.
For $n_{proc}=$ 40, however, the best speedup is for 4 OpenMP threads, dropping down when using 8 threads.

The overall speedup is dependent on the system and on the computers used, but in general is considered to be satisfactory allowing the use of around 100 cores for large calculations.

\section{Conclusions}
We have presented the MADWAVE3 program, a FORTRAN90 code devoted to the propagation of quantum wave packets for triatomic systems to compute state-to-state probabilities and cross sections of reactive and inelastic collisional (A+BC ($v,j$)$\rightarrow$ AB($v',j'$)+C) and photodissociation (ABC +h$\nu$ $\rightarrow$ A+ BC($v,j$)) processes.
The program can deal with several coupled electronic states in a diabatic representation.
These electronic states are coupled by potential terms, but not by non-adiabatic kinetic terms. The diabatic representation is considered to be diagonal on the reactants' side at long distances
while it can be coupled on the products' side.

Several auxiliary codes are distributed within the MADWAVE3 program to analyze and compute different magnitudes such as reactive or inelastic cross sections from the wave packet propagations.

The underlying theoretical background is explained, including the calculation of the reactants' and products' rovibrational states, construction of the initial wave packet, its propagation and analysis of the final wave packet to compute reaction probabilities.

The input structure of the code is presented, as FORTRAN namelists, and an exhaustive description of their variables is indicated in Table~\ref{tab:parameters}.

The D + H$_2^+$ $\rightarrow$ DH$^+$ + H is presented as an use example of the code and the results compared with those from ABC program.

The code is parallelized using MPI and OpenMP in the helicity $\Omega$ and the $\gamma$ angle between the two Jacobi vectors.

The most updated version of the MADWAVE3 code will always be available at \repourl. In its current form it requires the open source libraries for MPI and FFTW3. The source codes, the installation procedure and the example discussed in this work are also provided in the supplementary information as a zip file, in a format close to that found in the githup repository.

\section{Acknowledgments}

We express our gratitude to Carlos Mouri{\~ n}o, from CESGA supercomputer center in Galicia (Spain),
for his help in the parallelization of the code, specially
in during the implementation the OpenMP when using the non threads-safe library FFTW3.
This work has received funding from MICIN (Spain) under Grant Nos. PID2021-122549NB-C21, PID2021-122549NB-C22


\bibliographystyle{elsarticle-num}







\end{document}